\documentclass[useAMS]{mn2e}

\usepackage{graphicx}
\usepackage{epsf}
\usepackage{times}
\usepackage{color}

\newcommand{\HI}{{H}\,{\sc i}}
\newcommand{\Halpha}{H$\alpha$}

\newcommand{\etal}{{et al.}}
\newcommand{\kms}{km s$^{-1}$}
\newcommand{\e}[1]{$\,\pm\,$#1}

\def\mlstard{\ifmmode\Upsilon_\mathrm{d}\else$\Upsilon_\mathrm{d}$\fi}
\def\mlstarb{\ifmmode\Upsilon_\mathrm{b}\else$\Upsilon_\mathrm{b}$\fi}
\def\mlstarR{\ifmmode\Upsilon^R\else$\Upsilon^R$\fi}
\def\mlstarsR{\ifmmode\Upsilon_\ast^R\else$\Upsilon_\ast^R$\fi}
\def\mlstardR{\ifmmode\Upsilon^R_\mathrm{d}\else$\Upsilon^R_\mathrm{d}$\fi}
\def\mlstarbR{\ifmmode\Upsilon^R_\mathrm{b}\else$\Upsilon^R_\mathrm{b}$\fi}
\def\HIm{\ifmmode \hbox{\scriptsize H\kern0.5pt{\footnotesize\sc i}}\else
H\kern1pt{\small I}\fi}
\def\apj{ApJ}
\def\apjl{ApJ}
\def\aj{AJ}
\def\aap{A\&A}
\def\mnras{MNRAS}
\title[Baryon Distribution and Rotation Curve Shape]{The Link between the Baryonic Mass Distribution and the Rotation Curve Shape}

\author[Swaters, Sancisi, van der Hulst, \& van Albada]{R. A. Swaters,$^1$
\thanks{E-mail: rob@swaters.net}
R. Sancisi,$^{2,3}$ J. M. van der Hulst,$^3$ T. S. van Albada$^3$\\
$^1$ National Optical Astronomy Observatory, 950 North Cherry Ave.,
Tucson, AZ 85719\\
$^2$ INAF - Osservatorio Astronomico di Bologna, via Ranzani 1, 40127
Bologna, Italy \\
$^3$ Kapteyn Astronomical Institute, University of Groningen, Landleven 12, 
9747 AD Groningen, the Netherlands }

\begin{document}

\maketitle

\label{firstpage}

\begin{abstract}

  The observed rotation curves of disc galaxies, ranging from
  late-type dwarf galaxies to early-type spirals, can be fit
  remarkably well simply by scaling up the contributions of the
  stellar and \HI\ discs. This `baryonic scaling model' can explain
  the full breadth of observed rotation curves with only two free
  parameters. For a small fraction of galaxies, in particular
  early-type spiral galaxies, HI scaling appears to fail in the outer
  parts, possibly due to observational effects or ionization of the
  \HI.  The overall success of the baryonic scaling model suggests
  that the well-known global coupling between the baryonic mass of a
  galaxy and its rotation velocity (known as the baryonic Tully-Fisher
  relation), applies at a more local level as well, and it seems to
  imply a link between the baryonic mass distribution and the
  distribution of total mass (including dark matter).

\end{abstract}

\begin{keywords}
galaxies: dwarf -- galaxies: kinematics and dynamics -- Galaxies:
spiral.
\end{keywords}

\section{Introduction}
\label{theintro}

Over the bright optical part of spiral galaxy discs, observed
rotation curves can usually be explained by the contribution of the
stars alone (e.g., Kalnajs 1983; Kent 1986). In this so-called maximum
disc hypothesis, the contribution of stellar disc to the rotation
curve is maximized, and that of the dark matter is minimized (e.g.,
van Albada \etal\ 1985; Begeman 1987; Broeils 1992a). However,
even though the contribution of the stellar disc can be scaled to
explain the inner rotation curve, good fits can usually be obtained
for a range of contributions of the stellar discs, including
submaximal and minimal discs (e.g., van Albada \etal\ 1985; Verheijen
1997; Swaters 1999; Dutton \etal\ 2005; Noordermeer 2006).

In almost all galaxies the contribution of the stellar disc can be
scaled to explain all of the observed rotation curve out to about two
or three disc scale lengths (e.g., Kalnajs 1983; Kent 1986; Palunas \&
Williams 2000), regardless of the relative contribution of the stellar
disc to the rotation curve. This is not an argument that discs are
maximal, but it does indicate that the total mass density (which
includes the dark matter) and the luminous mass density are closely
coupled (e.g., Sancisi 2004; Swaters \etal\ 2009).

This link between the distributions of luminous mass and total mass is
not limited to spiral galaxies.  Recently, Swaters \etal\ (2011)
showed that for all but one of the late-type dwarf galaxies in their
sample, the contribution of the stellar disc to the rotation curve can
also be scaled up to explain most of the inner parts of the observed
rotation curves. Even though these galaxies may not actually have
maximal discs, because the required stellar mass-to-light ratios are
high, up to about 15 in the $R$-band, the fact that the stellar disc
can be scaled to explain the inner two or three discs scale lengths of
the rotation curves demonstrates that the apparent coupling between
luminous and total mass density extends to late-type dwarf galaxies as
well.

Interestingly, the apparent coupling between the rotation curve shape
and the visible matter works not only for the stellar disc, but also
for the \HI\ disc. Bosma (1978, 1981) found that the ratio of total
mass surface density to \HI\ surface density is roughly constant in
the outer parts of galaxies. He also found that the ratio decreased
towards galaxies with lower rotation velocities.  This proportionality
of \HI\ to dark mass surface density was later found in late-type
dwarf galaxies (Carignan 1985; Carignan \& Puche 1990; Jobin \&
Carignan 1990).

Hoekstra, van Albada \& Sancisi (2001; hereafter HvAS) tested the
hypothesis that the \HI\ disc can be scaled up to explain the outer
rotation curve for the large sample of galaxies presented in Broeils
(1992a), which spans a large range in galaxy properties.  HvAS found
that for most of the galaxies in this sample the rotation curves can
be explained well by scaling up the \HI\ contribution by a factor of
around 10, in combination with a maximal optical disc.  In contrast
with Bosma (1978), HvAS found that the \HI\ scaling factor was
independent of the maximum rotation velocity.  They noted that among
early-type spiral galaxies \HI\ scaling was less successful at
explaining the observed rotation curves.  They did not conclude that
there is a real coupling between HI and dark matter.

\begin{table*}
\begin{minipage}{\hsize}
\caption{Optical and dark matter properties\label{tabfits}}
\begin{center}
\begin{tabular}{lcrrrrrrrrrrrrr}
\hline
UGC & Source & T & $\mathrm{D}_a$ & $M_R$ & $h$ & $\mu_0^R$ & $r_\mathrm{lmp}/h$ & $v_\mathrm{lmp}$ & $i$ & $\mlstarR_\mathrm{d,MD}$ & $\chi^2_\mathrm{MD}$ & $\mlstarR_\mathrm{d,HI}$ & $\eta$ & $\chi^2_\mathrm{HI}$ \\
(1) & (2) & (3) & (4) & (5) & (6) & (7) & (8) & (9) & (10) & (11) & (12) & (13) & (14) & (15) \\
\hline
  731 & dwarf & 10 &  8.0 &-16.6 &1.7  &23.0 &4.2 & 74 &57 & 15.1 & 0.5 & 14.0\e1.2 &  3.8\e0.8 & 0.2 \\
 2916 & early &  2 & 63.5 &-22.0 &5.0  &21.0 & 7.4&181 &42 &  5.2 & 0.4 &  4.9\e1.0 &  5.2\e1.4 & 0.5 \\
 2953 & early &  2 & 15.1 &-22.5 &4.1  &19.3 &14.6&283 &50 &  5.4 & 2.0 &  5.1\e0.2 &   45\e3   & 3.0 \\
 3205 & early &  2 & 48.7 &-21.9 &3.5  &19.6 &11.1&219 &67 &  4.5 & 4.0 &  4.5\e0.2 & 10.6\e0.6 & 5.4 \\
 3371 & dwarf & 10 & 12.8 &-17.7 &3.1  &23.3 &3.3 & 86 &49 & 12.5 & 0.7 & 12.1\e1.0 &  4.9\e1.0 & 0.5 \\
 3546 & early &  1 & 27.3 &-21.4 &2.8  &19.5 &10.0&193 &55 &  4.0 & 1.0 &  4.6\e0.3 &   20\e3   & 1.4 \\
 3580 & early &  1 & 19.2 &-19.4 &2.4  &21.6 &10.5&124 &63 &  7.2 & 3.2 &  5.1\e0.5 & 10.9\e0.6 & 3.4 \\
 4325 & dwarf &  9 & 10.1 &-18.1 &1.6  &21.6 &3.6 & 92 &41 &  9.1 & 1.8 &  8.9\e0.8 &  0.8\e0.8 & 1.1 \\
 4499 & dwarf &  8 & 13.0 &-17.8 &1.5  &21.5 &5.7 & 74 &50 &  2.3 & 0.6 &  3.1\e0.5 &  5.3\e0.5 & 1.0 \\
 5414 & dwarf & 10 & 10.0 &-17.6 &1.5  &21.8 &2.9 & 61 &55 &  4.6 & 1.1 &  3.1\e0.9 &  3.9\e1.6 & 0.8 \\
 6399 & UMa   &  9 & 18.6 &-18.7 &2.2  &22.2 &3.7 & 88 &75 &  2.7 & 0.1 &  2.9\e0.6 &  6.4\e2.5 & 0.1 \\
 6446 & dwarf &  7 & 12.0 &-18.4 &1.9  &21.4 &5.1 & 80 &52 &  4.0 & 0.4 &  4.3\e0.3 &  5.0\e0.5 & 1.1 \\
 6446 & UMa   &  7 & 18.6 &-18.9 &3.1  &22.2 &5.1 & 80 &51 &  3.2 & 0.7 &  3.0\e0.2 &  3.2\e0.5 & 0.8 \\
 6537 & UMa   &  5 & 18.6 &-21.7 &5.2  &20.1 &6.5 &167 &53 &  1.7 & 2.3 &  1.7\e0.1 &  5.3\e0.9 &10.1 \\ 
 6595 & UMa   &  3 & 18.6 &-20.4 &1.8  &19.5 &21.0&113 &76 &  0.8 & 0.8 &  0.5\e0.2 & 12.9\e1.3 & 0.8 \\ 
 6745 & UMa   &  5 & 18.6 &-21.6 &3.0  &19.1 &3.9 &169 &76 &  0.6 & 1.3 &  1.8\e0.2 &  4.7\e3.4 & 2.0 \\ 
 6778 & UMa   &  5 & 18.6 &-18.7 &2.5  &19.3 &8.4 &148 &49 &  1.7 & 1.4 &  1.6\e0.2 & 10.9\e1.6 & 3.0 \\ 
 6786 & early & -1 & 25.9 &-21.1 &1.5  &19.3 &20.1&211 &68 &  3.6 & 1.7 &  7.6\e0.7 &   42\e3   & 4.8 \\
 6787 & early &  2 & 18.9 &-32.3 &3.3  &20.5 &10.1&255 &69 &  9.3 & 3.4 &  8.1\e0.2 &   52\e4   & 2.6 \\
 6815 & UMa   &  6 & 18.6 &-20.8 &2.9  &20.9 &5.3 &137 &79 &  1.6 & 2.0 &  2.3\e0.1 &  7.4\e1.3 & 2.8 \\ 
 6869 & UMa   &  4 & 18.6 &-21.0 &1.7  &19.2 &4.3 &169 &55 &  0.5 & 1.0 &  1.1\e0.1 & 10.1\e2.6 & 2.5 \\ 
 6870 & UMa   &  4 & 18.6 &-22.1 &3.9  &19.5 &4.2 &215 &62 &  2.3 & 0.6 &  2.6\e0.1 & 11.1\e3.0 & 1.0 \\ 
 6904 & UMa   &  4 & 18.6 &-20.2 &2.1  &19.9 &4.4 &134 &77 &  1.5 & 0.9 &  1.8\e0.2 & 12.6\e2.3 & 1.2 \\ 
 6917 & UMa   &  7 & 18.6 &-19.6 &3.4  &21.9 &3.2 &111 &56 &  3.2 & 0.6 &  3.3\e0.2 &  5.3\e1.2 & 0.5 \\
 6937 & UMa   &  4 & 18.6 &-22.2 &4.1  &19.2 &8.7 &237 &56 &  4.1 & 0.7 &  4.3\e0.3 & 16.0\e1.9 & 6.9 \\ 
 6983 & UMa   &  6 & 18.6 &-19.4 &3.6  &22.0 &4.5 &109 &49 &  4.7 & 0.8 &  4.6\e0.3 &  3.6\e0.8 & 1.0 \\
 7095 & UMa   &  4 & 18.6 &-21.4 &2.8  &19.2 &8.3 &159 &73 &  2.7 & 1.8 &  2.6\e0.1 & 13.1\e1.4 & 1.8 \\ 
 7323 & dwarf &  8 &  8.1 &-18.9 &2.2  &21.2 &2.7 & 86 &47 &  3.0 & 0.4 &  2.9\e0.3 &  4.9\e1.5 & 0.4 \\
 7399 & dwarf &  8 &  8.4 &-17.1 &0.79 &20.7 &13.9&109 &55 &  7.7 & 1.3 &  4.7\e0.5 & 21.3\e0.6 &12.0 \\
 7524 & dwarf &  9 &  3.5 &-18.1 &2.6  &22.2 &3.1 & 79 &46 &  7.2 & 0.3 &  6.9\e0.5 &  4.0\e0.8 & 0.2 \\
 7559 & dwarf & 10 &  3.2 &-13.7 &0.67 &23.8 &3.1 & 33 &61 & 13.1 & 0.2 & 10.8\e5.2 &  4.3\e2.5 & 0.2 \\
 7577 & dwarf & 10 &  3.5 &-15.6 &0.84 &22.5 &2.7 & 18 &63 &  0.9 & 0.4 &  0.7\e0.4 &  1.5\e1.5 & 0.3 \\
 7603 & dwarf &  7 &  6.8 &-16.9 &0.90 &20.8 &6.6 & 64 &78 &  4.1 & 0.8 &  2.7\e0.5 &  9.2\e0.9 & 1.7 \\
 8490 & dwarf &  9 &  4.9 &-17.3 &0.66 &20.5 &16.2& 78 &50 &  4.4 & 0.4 &  3.3\e0.4 & 12.2\e0.5 & 1.2 \\
 8699 & early &  2 & 36.7 &-20.7 &3.7  &22.2 &6.5 &183 &73 & 12.0 & 0.8 &  8.2\e1.1 & 17.3\e2.5 & 1.1 \\
 9133 & early &  2 & 54.3 &-22.6 &9.1  &21.3 &11.2&229 &53 & 10.0 & 1.0 &  8.1\e0.6 & 13.0\e1.3 & 1.6 \\
 9211 & dwarf & 10 & 12.6 &-16.2 &1.3  &22.6 &6.3 & 65 &44 & 11.2 & 0.6 & 10.1\e1.8 &  6.2\e0.9 & 0.2 \\
11670 & early &  0 & 12.7 &-20.6 &1.8  &19.6 &13.3&171 &70 &  3.7 & 5.6 &  2.9\e0.2 &   37\e3   & 9.4 \\
11707 & dwarf &  8 & 15.9 &-18.6 &4.3  &23.1 &3.5 &100 &68 &  9.3 & 0.8 &  9.4\e0.8 &  2.9\e0.4 & 2.0 \\
11852 & early &  1 & 80.0 &-21.5 &4.5  &20.7 &20.6&165 &50 &  8.5 & 0.3 &  7.6\e0.9 &  7.7\e0.8 & 0.4 \\
12043 & early &  0 & 15.4 &-18.6 &0.84 &19.9 &19.6& 94 &67 &  2.6 &10.2 &  1.9\e0.2 & 14.4\e0.4 & 6.6 \\
12060 & dwarf & 10 & 15.7 &-17.9 &1.8  &21.6 &5.8 & 74 &40 &  8.3 & 0.2 &  7.9\e0.8 &  3.2\e1.2 & 0.2 \\
12632 & dwarf &  9 &  6.9 &-17.1 &2.6  &23.5 &3.3 & 76 &46 & 15.1 & 1.1 & 14.1\e1.2 &  3.7\e0.9 & 1.3 \\
12732 & dwarf &  9 & 13.2 &-18.0 &2.2  &22.4 &7.0 & 98 &39 &  7.5 & 0.4 &  8.0\e0.6 &  5.7\e0.5 & 0.8 \\
\hline
\end{tabular}
\end{center}
\hbox to\hsize{\hfil\vtop{\hsize=158mm 
  {(1) UGC number, (2) source of the data; for the dwarf
  galaxies, the rotation curves and inclinations come from Swaters
  \etal\ (2009), distances and photometry from Swaters \& Balcells
  (2002); for the Ursa Major cluster galaxies, all data come from
  Verheijen (1997), converted to the adopted cluster distance of
  18.6~Mpc (Tully \& Pierce 2000); for the early-type galaxies, data
  come from Noordermeer (2006) and Noordermeer \etal\ (2007), (3)
  numeric Hubble type, (4) adopted distance in Mpc, (5) absolute
  $R$-band magnitude, (6) $R$-band scale length in kpc, (7)
  extrapolated central $R$-band disc surface brightness , (8) radial
  extent of the rotation curve in units of scale lengths, (9) the
  rotation velocity at the last measured point in \kms, (10)
  inclination in degrees, (11, 12) maximum disc fit mass-to-light
  ratio, in units of M/L$_\odot$, and reduced $\chi^2$ value for the
  fit (see Verheijen 1997; Noordermeer 2006; Swaters \etal\ 2011),
  (13,14,15) best-fitting values for the mass-to-light ratio, \HI\ scale
  factor $\eta$, and reduced $\chi^2$ value for the \HI\ scaling
  model.}}\hfil}
\end{minipage}
\end{table*}

Hessman \& Ziebart (2011) revisited this hypothesis, which they dubbed
`the Bosma effect', using a subset of 17 galaxies from The \HI\ Nearby
Galaxy Survey of de Blok \etal\ (2008). They concluded that the
observed rotation curves could be explained nearly as well by scaling
up the \HI\ and stellar discs as by their models based on Burkert and
NFW haloes.

The reason why the HI scaling appears to work and can reproduce the
observed rotation curves is not clear. There have been various
suggestions, including the presence of large amounts of cold gas in
the disc (Pfenniger \& Combes 1994), and hybrid models with both dark
matter and cold gas (Revaz \etal\ 2009). It is important to emphasize
here that the success of the \HI\ scaling does not mean that all mass
must reside in the baryonic disc, nor that dark matter haloes are
unnecessary.

\begin{figure*}
\includegraphics[width=178mm]{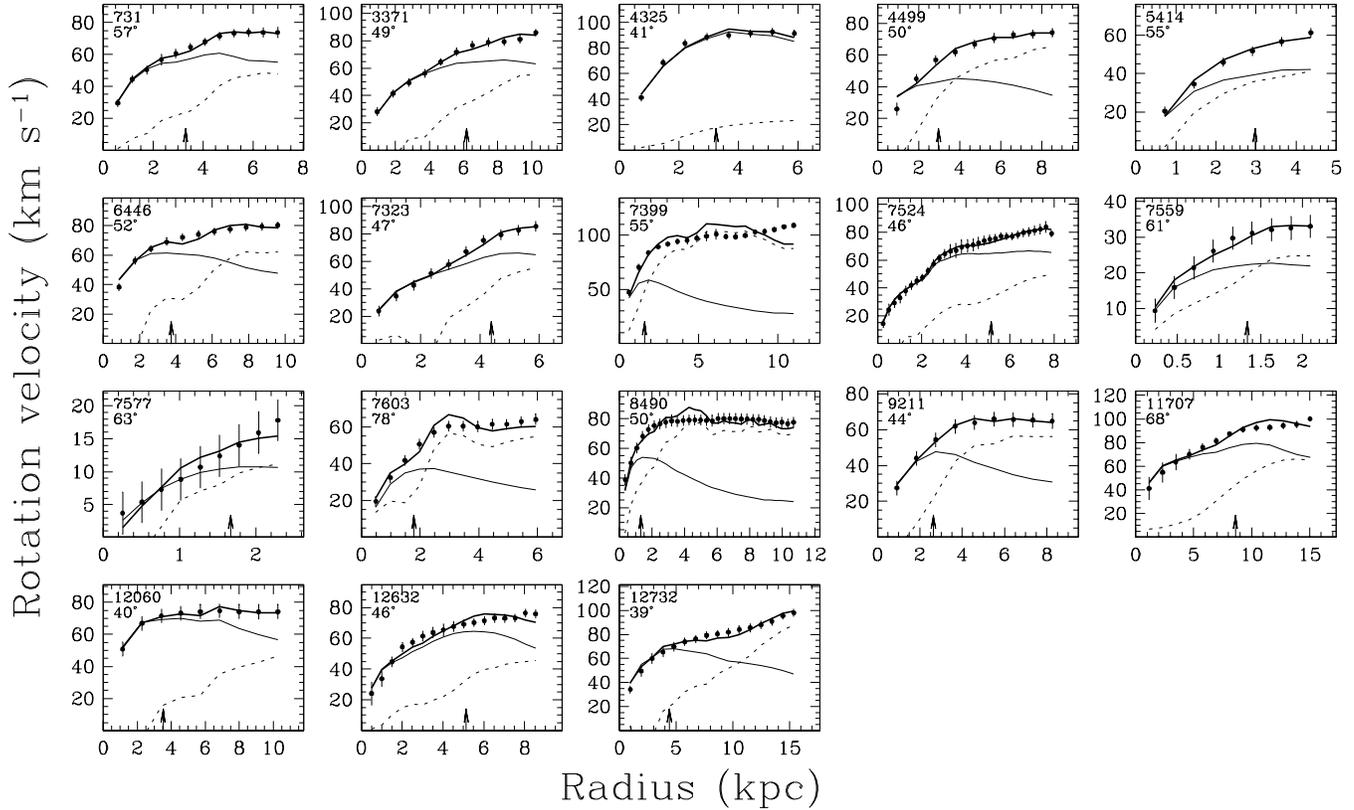}
\caption{ Mass models based on scaling the stellar disc and the
  \HI\ component for the late-type dwarf galaxies in our sample. The
  filled circles represent the derived rotation curves.  The thin full
  lines represent the contribution of the stellar discs to the
  rotation curves and the dotted lines that of the gas. The thick
  solid lines represent the best-fitting model based on scaling the
  contributions of the stars and the gas. The arrows at the bottom of
  each panel indicate a radius of two optical disc scale lengths.  In
  the top left corner of each panel the UGC number and the inclination
  are given.}
\label{figdwarfs}
\end{figure*}

The purpose of the present article is to further explore how well the
\HI\ scaling works and to extend the study of HvAS to the larger
sample of high-quality rotation curves that have recently become
available for late-type dwarf galaxies (Swaters 1999; Swaters
\etal\ 2009), early-type spiral galaxies (Noordermeer 2006), and
late-type spiral galaxies in the Ursa Major cluster (Verheijen 1997;
Verheijen \& Sancisi 2001).

\section{The sample and the rotation curves}
\label{thesample}

The sample presented here was selected from the dwarf galaxies
presented in Swaters \etal\ (2009), from the Ursa Major sample from
Verheijen (1997), which contains a large fraction of late-type spiral
galaxies, and the early-type spiral galaxies presented in Noordermeer
(2006).

The sample of late-type dwarf galaxies used here was observed as part
of the WHISP project (Westerbork HI Survey of Spiral and Irregular
Galaxies; see van der Hulst, van Albada \& Sancisi (2001); Swaters
\etal\ 2002).  The galaxies in the WHISP sample were selected from the
UGC catalogue (Nilson 1973), taking all galaxies with declinations
north of $20^\circ$, blue major axis diameters larger than $1.5'$ and
measured flux densities larger than 100 mJy. Selected from this list
were the late-type dwarf galaxies, defined as galaxies with Hubble
types later than Sd, supplemented with spiral galaxies of earlier
Hubble types but with absolute $B$-band magnitudes fainter than $-17$.

The galaxies in this sample have inclinations in the range
$39^\circ\le i<80^\circ$.  The lower limit of $39^\circ$ was chosen to
include UGC~12732 as well. Most of the galaxies in this sample have
absolute magnitudes fainter than $M_R=-18$, as expected for dwarf
galaxies. Only the 18 galaxies with the highest quality rotation
curves as assigned by Swaters \etal\ (2009) were selected.  The
rotation curves were derived with a procedure that corrects for the
effects of beam smearing to a large degree (Swaters \etal\ 2009).
More details about this sample and its selection are provided in
Swaters \etal\ (2011).

\begin{figure*}
\includegraphics[width=178mm]{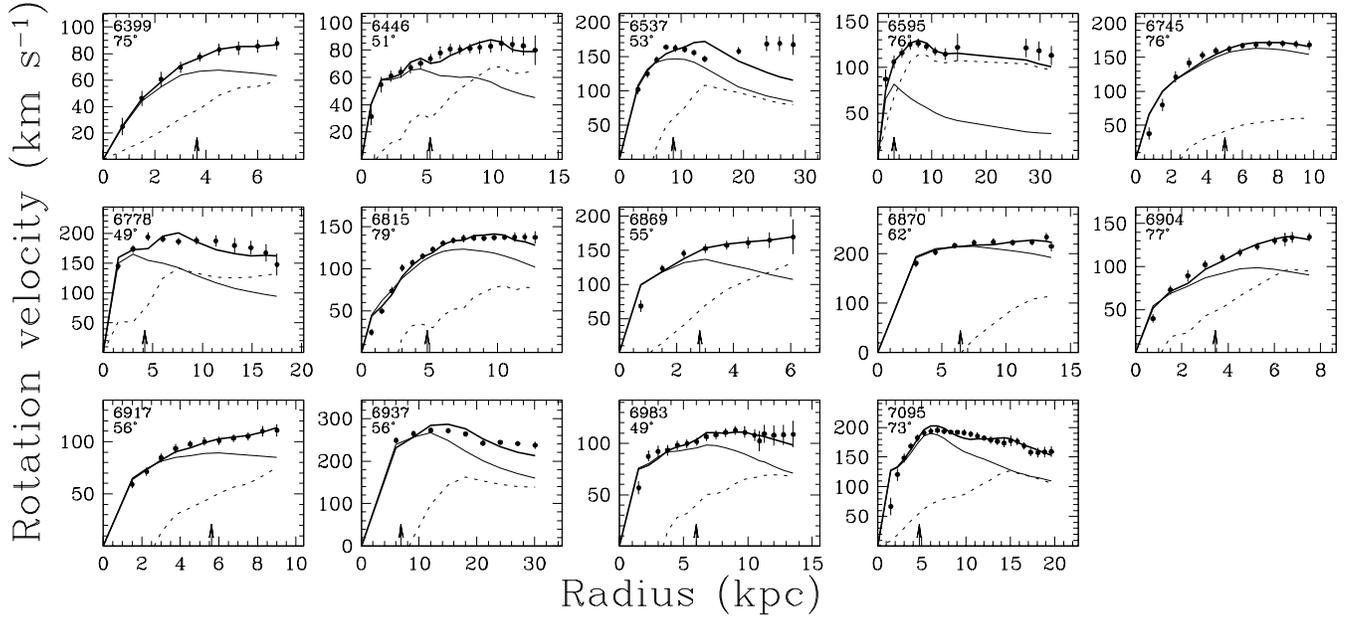}
\caption{ Same as Fig. \ref{figdwarfs} but for the galaxies in the
    Ursa Major sample.}
\label{figUMa}
\end{figure*}

The late-type spiral galaxies in this sample used in this paper come
from the \HI\ study of galaxies in the Ursa Major cluster by Verheijen
(1997; see also Verheijen \& Sancisi 2001 and Verheijen 2001). This
cluster is dominated by late-type spiral galaxies, but also contains
galaxies with earlier and later types. The cluster is assumed to be at
a distance of 18.6 Mpc (Tully \& Pierce 2000).

We selected the galaxies from the sample presented in Verheijen
(2001), which only includes galaxies with inclinations larger than
$45^\circ$. In addition, we assigned a quality assessment to each
rotation curve following Swaters \etal\ (2009), and only selected the
14 galaxies with the highest quality.  The rotation curves in the Ursa
Major sample have been corrected for the effects of beam smearing,
through visual inspection and correction (see Verheijen \& Sancisi
2001).

Finally, the early-type spiral galaxies come from the sample of
Noordermeer (2006), also selected from the WHISP survey. For the
selection of the early-type galaxies, however, a lower flux limit of
20 mJy was used, made possible because of improvements to the
Westerbork Synthesis Radio Telescope. More details about the sample
selection can be found in Noordermeer \etal\ (2005).

The rotation curves for this sample are presented in Noordermeer
\etal\ (2007). For this sample, the effects of beam smearing were
mitigated by the use of high-resolution \Halpha\ long-slit
spectroscopic observations. For the study presented here, we only
selected the 12 galaxies that have inclinations between
$40^\circ$ and $80^\circ$ degrees, excluding the lowest quality
rotation curves as given in Noordermeer \etal\ (2007).

We note that one galaxy, UGC 6446, is included in both the late-type
dwarf and the Ursa Major sample. Because the data and the derived
properties are completely independent, we have included both as an
independent illustration of the uncertainties on the derived
properties.

\section{Mass models}
\label{themodels}

Assuming that a galaxy is axially symmetric and in equilibrium, its
gravitational potential can be expressed as the sum of the individual
mass components: 
\begin{equation}
{v_\mathrm{c}}^2 = {v_\mathrm{*}}^2 +
{v_\mathrm{g}}^2 + {v_\mathrm{h}}^2.
\label{eqprefit}
\end{equation}
Here, $v_\mathrm{c}$ is the circular velocity, which is directly
linked to the gravitational potential.  The contribution of the stars
to the rotation curve is given by $v_\mathrm{*}$, $v_\mathrm{g}$
represents the contribution of the gas, and $v_\mathrm{h}$ that of the
dark halo.

\begin{figure*}
\includegraphics[width=178mm]{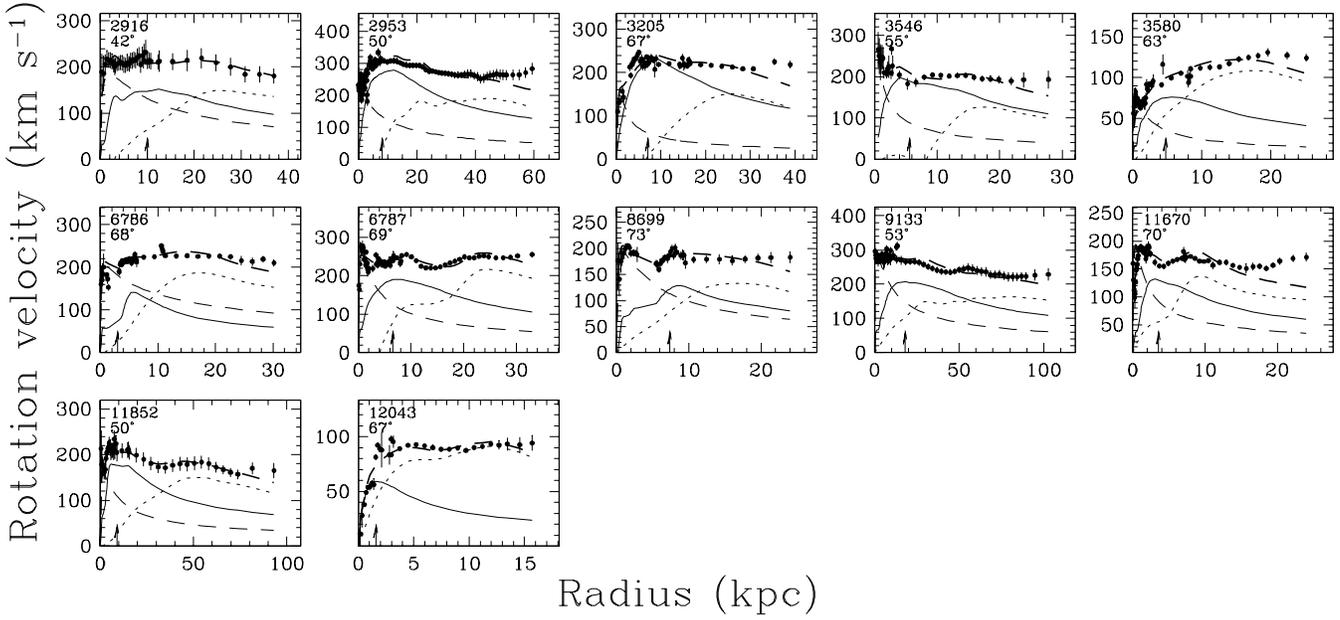}
\caption{ Same as Fig. \ref{figdwarfs} but for the early-type
  galaxies. The long-dashed line represents the contribution of the
  stellar bulge. }
\label{figearly}
\end{figure*}

The contribution of the stars to the rotation curve consists of a
bulge (usually not present in dwarf galaxies and late-type spiral
galaxies) and the disc. Because we assume no prior knowledge of the
stellar mass-to-light ratios, some value has to be assumed for both
the bulge (\mlstarb) and the disc (\mlstard). The contribution of the
neutral \HI\ gas to the rotation curve needs to be corrected for the
contribution of helium and other elements. However, because in this
paper we investigate how well the \HI\ component can be scaled to
explain the observed rotation curve, we use a free scale factor
$\eta$, omitting the dark halo term in Equation~\ref{eqprefit}. Making
this explicit, Equation~\ref{eqprefit} becomes:
\begin{equation}
{v_\mathrm{rot}} = \sqrt{\mlstarb{v_\mathrm{b}}^2 + \mlstard{v_\mathrm{d}}^2 +
\eta{v_\mathrm{\HIm}}^2},
\label{eqfit}
\end{equation}
where $v_\mathrm{b}$ and $v_\mathrm{d}$ are the contribution of the
bulge and stellar disc for a stellar mass-to-light ratio of unity, and
$v_\mathrm{\HIm}$ is that of the \HI\ only. This approach allows for
dark matter as well, but it requires the rotation curve of dark matter
is a linear combination of the rotation curves of the visible
components.

The rotation curves have not been corrected for asymmetric drift.  For
the late-type dwarf galaxies, as was described in more detail in
Swaters \etal\ (2009), the expected differences between the observed
rotation velocity and the circular velocity are small (less than 1
\kms\ in the inner parts of the rotation curves and less than 3 \kms\
at all radii for 95\% of the galaxies presented here). The spiral
galaxies in our sample have higher rotation velocities, and asymmetric
drift is expected to be less important there. A similar conclusion was
reached by Dalcanton \& Stilp (2010), who found that asymmetric drift
is unlikely to be important for galaxies with rotation velocities over
75~\kms. We therefore used the observed rotation curves to represent
the circular velocities.

In this paper, we only present mass models based on \HI\ scaling. The
maximum disc fit results presented in Table~\ref{tabfits} are taken
from Swaters \etal\ (2011), Verheijen (1997), and Noordermeer (2006)
for the late-type dwarfs, late-type spirals, and early-type spirals,
respectively.

\subsection{The contribution of stars and gas}

The contribution of the stellar disc to the rotation curve was
calculated using the prescription given in Casertano (1983), using the
$R$-band luminosity profiles presented in Swaters \& Balcells (2002)
for the late-type dwarf galaxies, the $K$-band profiles presented in
Tully \etal\ (1996) for the galaxies in the Ursa Major sample, and the
$R$-band profiles from Noordermeer \& van der Hulst (2007) for the
early-type spiral galaxies.

The contribution of the HI was also calculated using the prescription
from Casertano (1983), using the \HI\ radial profiles from Swaters
\etal\ (2002), Verheijen \& Sancisi (2001), and Noordermeer
\etal\ (2005).

For the mass models presented here we have used the same assumptions
for the disc and bulge parameters as the original papers. This means
that there are small differences in the adopted thicknesses and
vertical density profiles. For example, both Noordermeer (2006) and
Swaters \etal\ (2011) adopted an exponential vertical distribution
with a scale height one fifth of the scale length. Verheijen (1997),
on the other hand, adopted a vertical sech$^2$ distribution, but with
the same relative scale height. For the \HI\ disc, both Swaters
\etal\ (2011) and Noordermeer (2006) assume the same vertical
distribution for the \HI\ as for the stellar disc, whereas Verheijen
(1997) adopted an infinitely thin \HI\ disc. However, these
differences are of little consequence for the study presented
here. Not only are the effects of these different assumptions less
than a few percent (Swaters 1999; Noordermeer 2006; Swaters
\etal\ 2011), but the effects are most important in the inner
regions. Thus, the assumptions will have no significant effect on the
derived \HI\ scale factor $\eta$.

For the bulge contribution, we used the same decomposition as
presented in Noordermeer (2006). Following Verheijen (1997), no
bulge-disc decomposition was used for the galaxies in the Ursa Major
sample, because most of the galaxies have weak or no bulges.

We assumed that the disc mass-to-light ratio \mlstardR\ and the bulge
mass-to-light ratio \mlstarbR\ are independent of radius.

\subsection{Rotation curve fits}

The relative contribution of each of the different mass components is
determined by a simultaneous fit of the right-hand side of
Eq.~\ref{eqfit} to the observed rotation curve. For dwarf and
late-type spiral galaxies, which have no or no significant bulge, this
means there are only two free parameters, \mlstardR\ and $\eta$. For
the early-type galaxies, \mlstarbR\ is the third free parameter.

All the fits to rotation curves are shown in Figs.~\ref{figdwarfs},
\ref{figUMa}, and \ref{figearly}. The fit parameters are listed in
Table~\ref{tabfits}. The uncertainties on the fitted parameters have
been derived from the 68\% confidence levels and include the
covariance between \mlstardR\ and $\eta$.  The uncertainties on the
derived rotation velocities are non-Gaussian, as the points in the
rotation curves are correlated, and the rotation curves and their
uncertainties can be affected by systematic effects. The confidence
levels and the corresponding uncertainties should be considered
estimates.

The bulge mass-to-light ratios \mlstarbR\ are not given in
Table~\ref{tabfits} because they are not used in this paper. However,
they are presented in Noordermeer (2006). The fits for the Ursa Major
galaxies where done with $K$-band light profiles presented in Tully
\etal\ (1996). For ease of comparison to the late-type dwarf and
early-type spiral galaxies, the $K$-band mass-to-light ratios were
converted to the $R$-band.

\section{Results}
\label{secresults}

As can be seen from Figures~\ref{figdwarfs}, \ref{figUMa}, and
\ref{figearly}, for the majority of the galaxies in our sample good
fits to the rotation curves can be obtained solely by scaling the
contributions of the stellar and \HI\ discs.  This is remarkable given
that there are only two free parameters in most of these fits, and
three free parameters for the early-type galaxies with
bulges. Even with only two or three free parameters, the \HI\
scaling model can explain rising rotation curves such as seen in
UGC~12732 (see Figure~\ref{figdwarfs}), and declining rotation curves
such as seen in UGC~7095 (see Figure~\ref{figUMa}). In some cases, it
can even explain large-scale features in the rotation curves,
such as UGC~6787 and UGC~11852 (see Figure~\ref{figearly}).

Although the model is successful at explaining the rotation
curves of the vast majority of the galaxies in our sample, there are
some problem cases.  For the late-type dwarf galaxies
(Figure~\ref{figdwarfs}), there is a clear discrepancy between the
model and the observed rotation curve for only one of the galaxies,
UGC~7399. We note that this galaxy has a lopsided appearance in \HI,
with the \HI\ extending almost twice further on one side of the galaxy
than on the other.  Because of this strong deviation from axisymmetry,
the derived \HI\ profile is likely not to be representative of the
true radial \HI\ distribution. In two other galaxies, UGC~7603 and
UGC~8490, the model does predict the overall shape correctly, but
there are some wiggles in the model that are not seen in the observed
rotation curve.

For the late-type spiral galaxies (Figure~\ref{figUMa}), there is also
only one galaxy for which \HI\ scaling clearly cannot explain the
observed rotation curve, UGC~6537. This galaxy has the strongest warp
in the Ursa Major sample, and for the most discrepant points in the
fit, the rotation curve has only been derived from one side.

For the early-type spiral galaxies (Figure~\ref{figearly}), the
\HI\ scaling does not appear to work as well as for the late-type
dwarf and late-type spiral galaxies. As was also noted by Noordermeer
(2006), the \HI\ scaling tends to give poorer results in the outer
regions (see e.g., UGC~2953, UGC~3205, UGC~3580, and
UGC~11670).

\begin{figure}
\includegraphics[width=84mm]{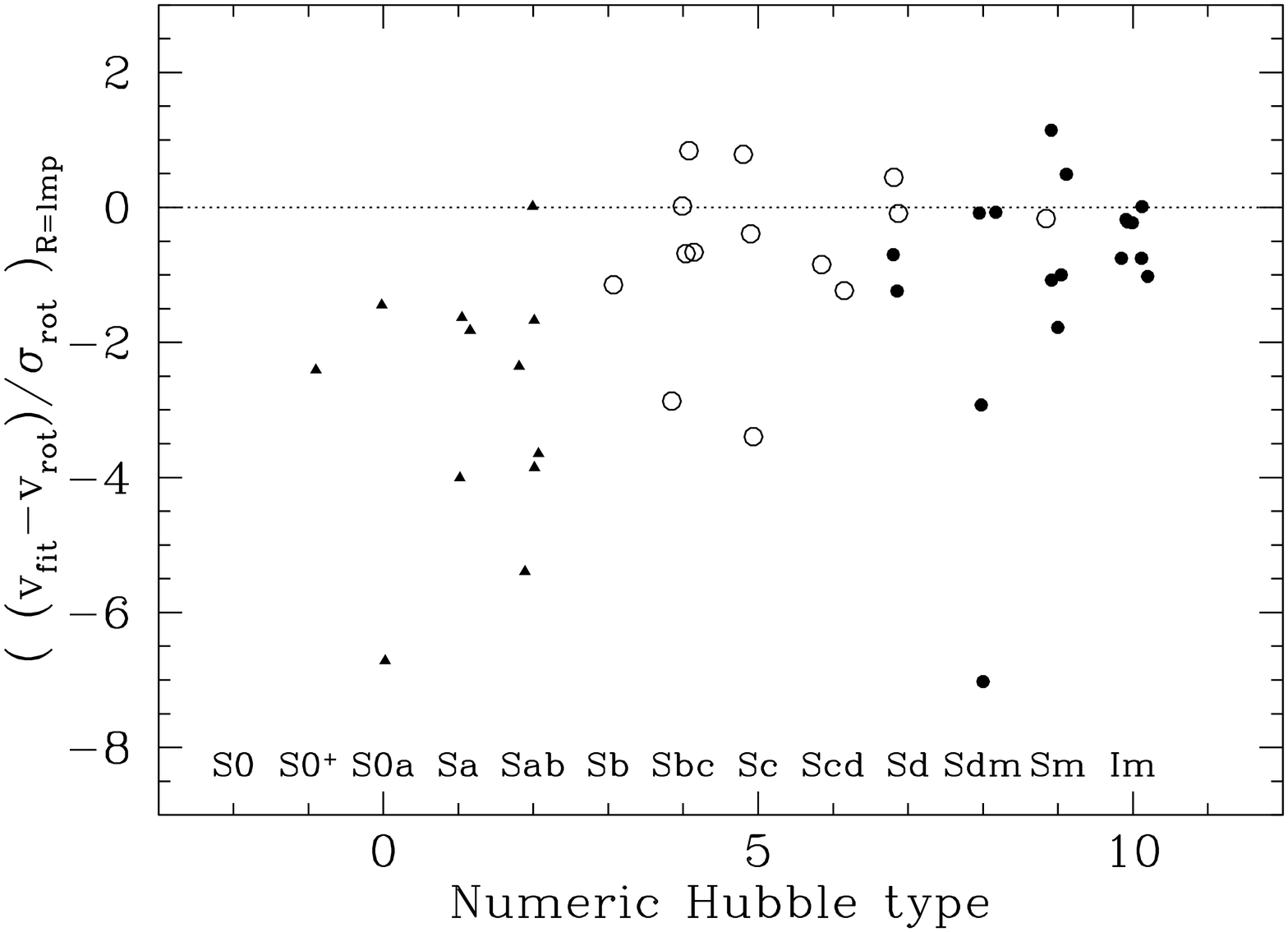}
\caption{ The difference between the fitted model and the observed
  rotation curve at the last measured point of the rotation curve,
  divided by the uncertainty on the rotation velocity. Triangles
  represent data from Noordermeer \etal\ (2007), circles represent
  data from Verheijen (2001), and dots represent data from Swaters
  \etal\ (2011). A random value between -0.4 and 0.4 has been added to
  the numeric Hubble type to spread the points in the plot. }
\label{figlmp}
\end{figure}

How well \HI\ scaling can explain the observed rotation curve at large
radii is shown in Figure~\ref{figlmp}, where the difference between
the model and the observed rotation velocity, divided by the
uncertainty in the rotation velocity, at the last measured point of
the rotation curve is plotted against the numeric Hubble type. Whereas
for late-type spiral and dwarf galaxies the model is close to observed
rotation curve (with the exception of UGC~7399 as explained above),
for early-type spiral galaxies the last measured point is on average
$3\sigma$ above the model.

\begin{figure}
\includegraphics[width=84mm]{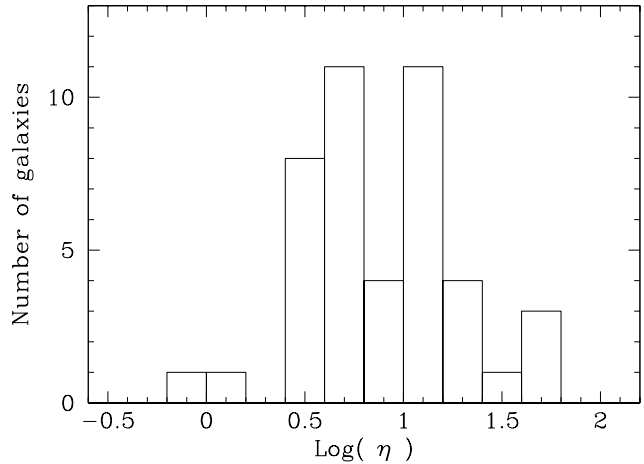}
\caption{ Distribution of the best-fitting \HI\ scaling factors $\eta$. }
\label{fighistmg}
\end{figure}

Interestingly, except perhaps for UGC~11670 and UGC~2953, the
discrepancy between our model and the observed rotation curves is seen
only in the data points derived from the velocity fields convolved to
$60''$. Although such a convolution is needed to reach the lowest
\HI\ column densities, the additional convolution will also exacerbate
the effects of beam smearing, especially at higher inclinations.
Indeed, for the seven galaxies with inclinations near $70^\circ$, the
model falls on average $4\sigma$ below the model, for the four
galaxies with inclinations near $50^\circ$ the difference is $2\sigma$
on average, and for the one galaxy near $40^\circ$ the model is in
agreement with the observed rotation curve. Although the relatively
small number of galaxies precludes a clear conclusion, such a
dependence on inclination is not expected if the model itself were
incorrect. We will discuss the early-type galaxies further in
Section~\ref{secdisc}.

\subsection{HI scale factors}
\label{sechiscale}

\begin{figure}
\includegraphics[width=84mm]{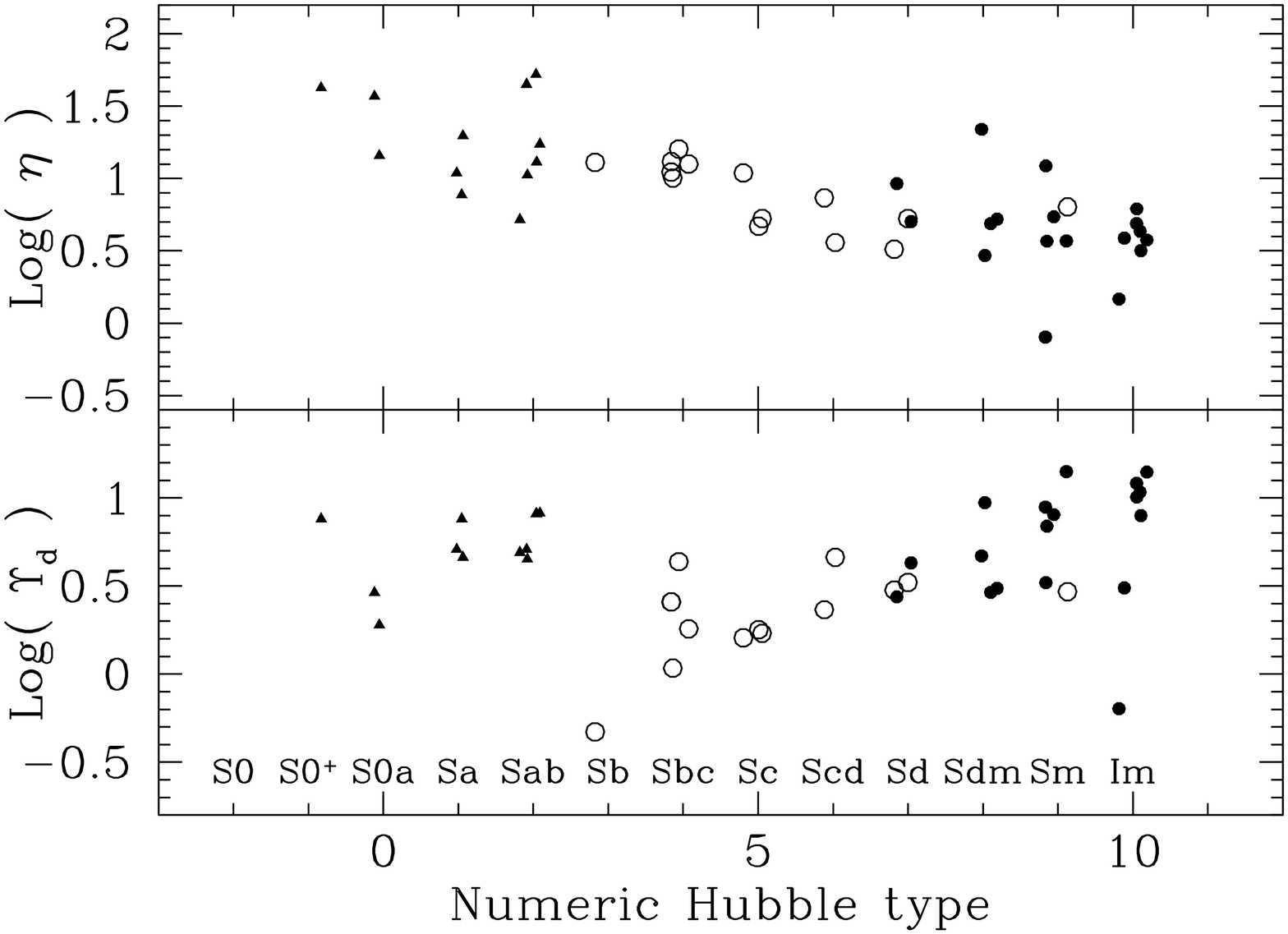}
\caption{ \HI\ scaling factors $\eta$ (top panel) and
    \mlstardR\ (bottom panel) vs. numerical Hubble type.  Symbol
  coding as in Figure~\ref{figlmp}. }
\label{figmgvsT}

\includegraphics[width=84mm]{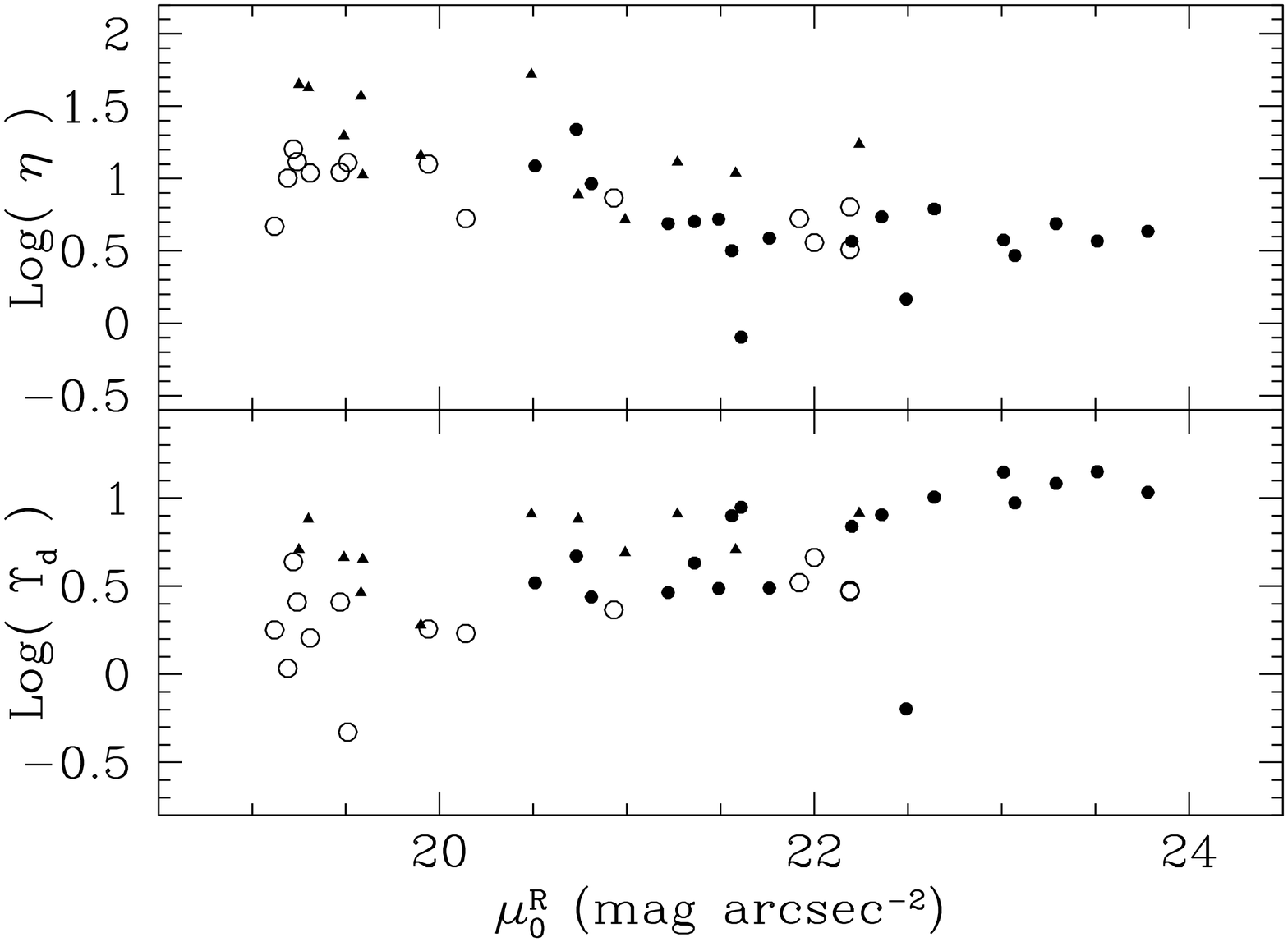}
\caption{ \HI\ scaling factors $\eta$ (top panel) and
    \mlstardR\ (bottom panel) vs. $R$-band extrapolated
  central disc surface brightness. Symbol coding as in Figure~\ref{figlmp}. }
\label{figmgvsmu}
\end{figure}

\begin{figure}
\includegraphics[width=84mm]{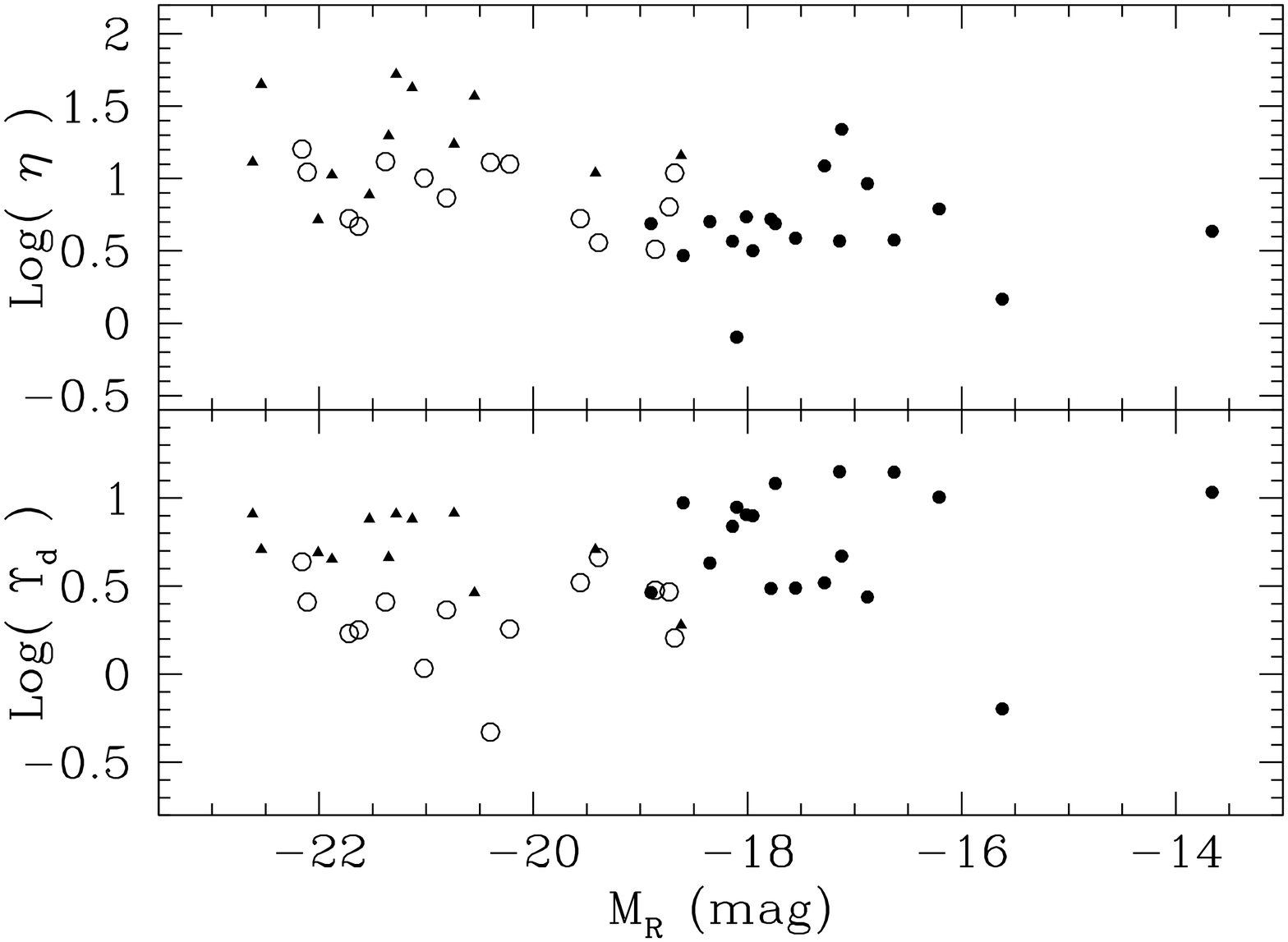}
\caption{ \HI\ scaling factors $\eta$ (top panel) and
    \mlstardR\ (bottom panel) vs. absolute $R$-band
  magnitude. Symbol coding as in Figure~\ref{figlmp}. }
\label{figmgvsM}

\includegraphics[width=84mm]{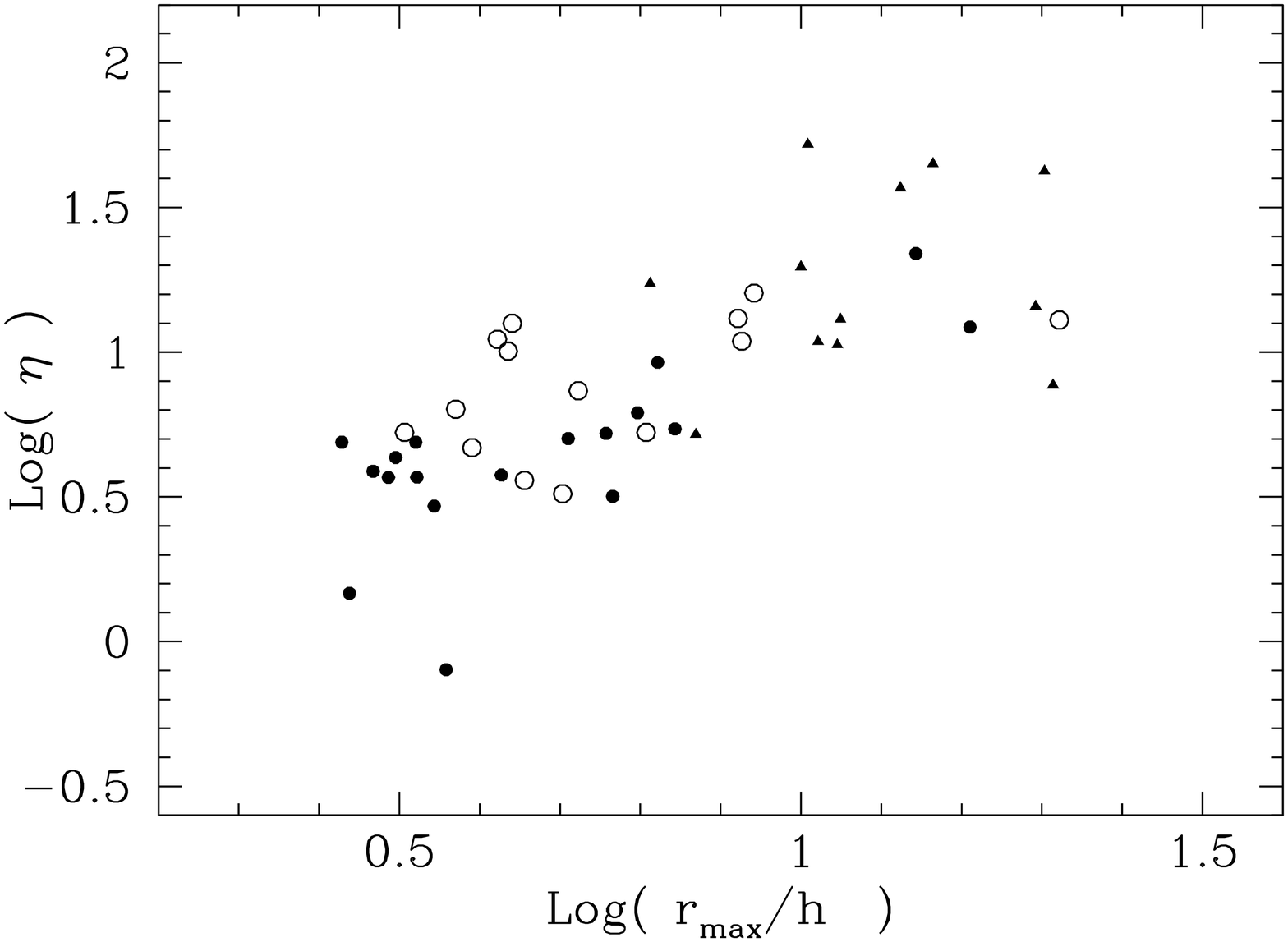}
\caption{ \HI\ scaling factors $\eta$ vs. maximum radial extent of the
  rotation curves in units of optical disc scale
  lengths. Symbol coding as in Figure~\ref{figlmp}. }
\label{figmgvshmax}
\end{figure}

Over our entire sample, the log-average scale factor is 8.2, similar
to the log-average of 10.2 found by HvAS. However, as can be seen in
Figure~\ref{fighistmg}, there is a considerable range in the scale
factors, ranging from near unity for two of the late-type dwarf
galaxies, to around fifty for some of the early-type spiral galaxies.

As can be seen in Figures~\ref{figmgvsT}, \ref{figmgvsmu}, and
\ref{figmgvsM}, the scale factor $\eta$ is correlated with numeric
Hubble type, central disc surface brightness, and absolute magnitude,
respectively. Galaxies with later Hubble types, fainter surface
brightnesses, and fainter absolute magnitudes tend to have lower scale
factors. Galaxies with a numeric Hubble type of 5 or later have a
log-average scale factor of $\sim 5$, whereas galaxies with earlier
Hubble types have a log-average $\eta$ of $\sim 16$.  Although not
shown here, $\eta$ is also correlated with the maximum rotation
velocity, as was already noted by Bosma (1978).

Interestingly, $\eta$ is also correlated with the extent of the
rotation curve, when expressed in units of disc scale lengths, as can
be seen in Figure~\ref{figmgvshmax}. Several factors, each discussed
below, likely play a role here, including the ratio of optical to
\HI\ scale lengths, the details of the fitting process, and intrinsic
correlations.

As was pointed out by HvAS, the extent of the rotation curve when
expressed in units of the \HI\ disc scale length (assuming that the
\HI\ distribution is roughly exponential) is relatively
constant. Ignoring possible central depressions, the central
\HI\ surface density is relatively constant among galaxies (about 7
M$_\odot$ pc$^{-2}$ for the galaxies in our sample with a dispersion
of 3 M$_\odot$ pc$^{-2}$). For the late-type spiral and dwarf galaxies
in our sample, the rotation curves could be derived out to an \HI\
surface density of about 1~M$_\odot$ pc$^{-2}$, corresponding to about
2 \HI\ scale lengths. For the early type galaxies, the \HI\
observations are more sensitive, and the rotation curve could
typically be determined out to an \HI\ surface density of about
0.25~M$_\odot$ pc$^{-2}$, corresponding to about three \HI\ scale
lengths.

This relatively constant extent of the rotation curves of roughly 2-3
\HI\ scale lengths means that if a rotation curve has a large ratio of
$r_\mathrm{max}/h$, the \HI\ is more extended relative to the optical
disc, and hence the rotation curve of the \HI\ only peaks farther
outside the optical disc. The more extended the \HI\ is relative to
the optical, the less the optical contributes at the radius where the
\HI\ peaks, and thus the \HI\ needs to be scaled up more to
explain the observed rotation curve.

Similarly, for early type galaxies, the presence of a maximum bulge
tends to depress the contribution of the stellar disc in the fits, and
thereby increases the scaling of the \HI\ component further. Also, if
the contribution of the stellar disc in the fits is not maximal (as
appears to be the case for UGC~7399 and UGC~8490 in
Figure~\ref{figdwarfs}, this also will result in a higher value for
$\eta$ in the fits.

Distance uncertainties can also affect the scale factors, but, given
the generally large distances to the early type galaxies, this will
not change the scale factors significantly.

However, all these effects are relatively small, and cannot explain
the high scale factors of up to 50 seen among some of the early type
galaxies.  Thus, it appears likely that some of the variation in scale
factors between galaxies is real.

\subsection{Stellar mass-to-light ratios}
\label{secmls}

\begin{figure}
\includegraphics[width=84mm]{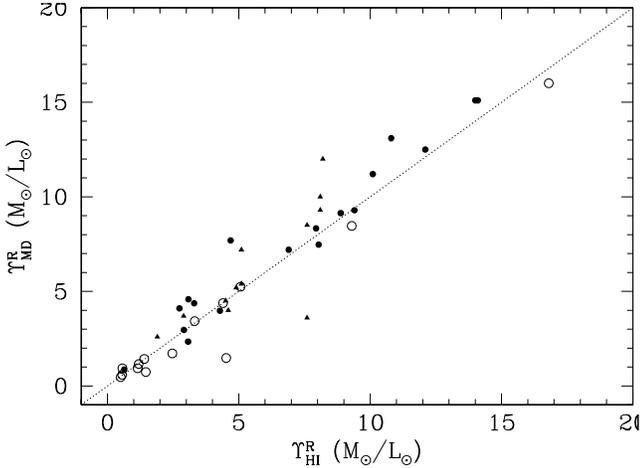}
\caption{ Comparison of the R-band mass-to-light ratios for the HI
  scaling models and those found in the maximum-disc fits. Symbol
  coding as in Figure~\ref{figlmp}. The dotted line is the line of
  equality. }
\label{figmaxdisc}
\end{figure}

As can be seen in Figure~\ref{figmaxdisc}, the stellar mass-to-light
ratios found in our models are similar to those found in the maximum
disc fits (see e.g., Verheijen 1997; Noordermeer 2006; Swaters
\etal\ 2011).  The values for \mlstardR\ range from 0.5 up to around
15. The high end of this range is well outside of the range predicted
by current stellar population synthesis models (e.g., Bell
\etal\ 2003; Zibetti, Charlot \& Rix 2009).

\begin{figure}
\includegraphics[width=84mm]{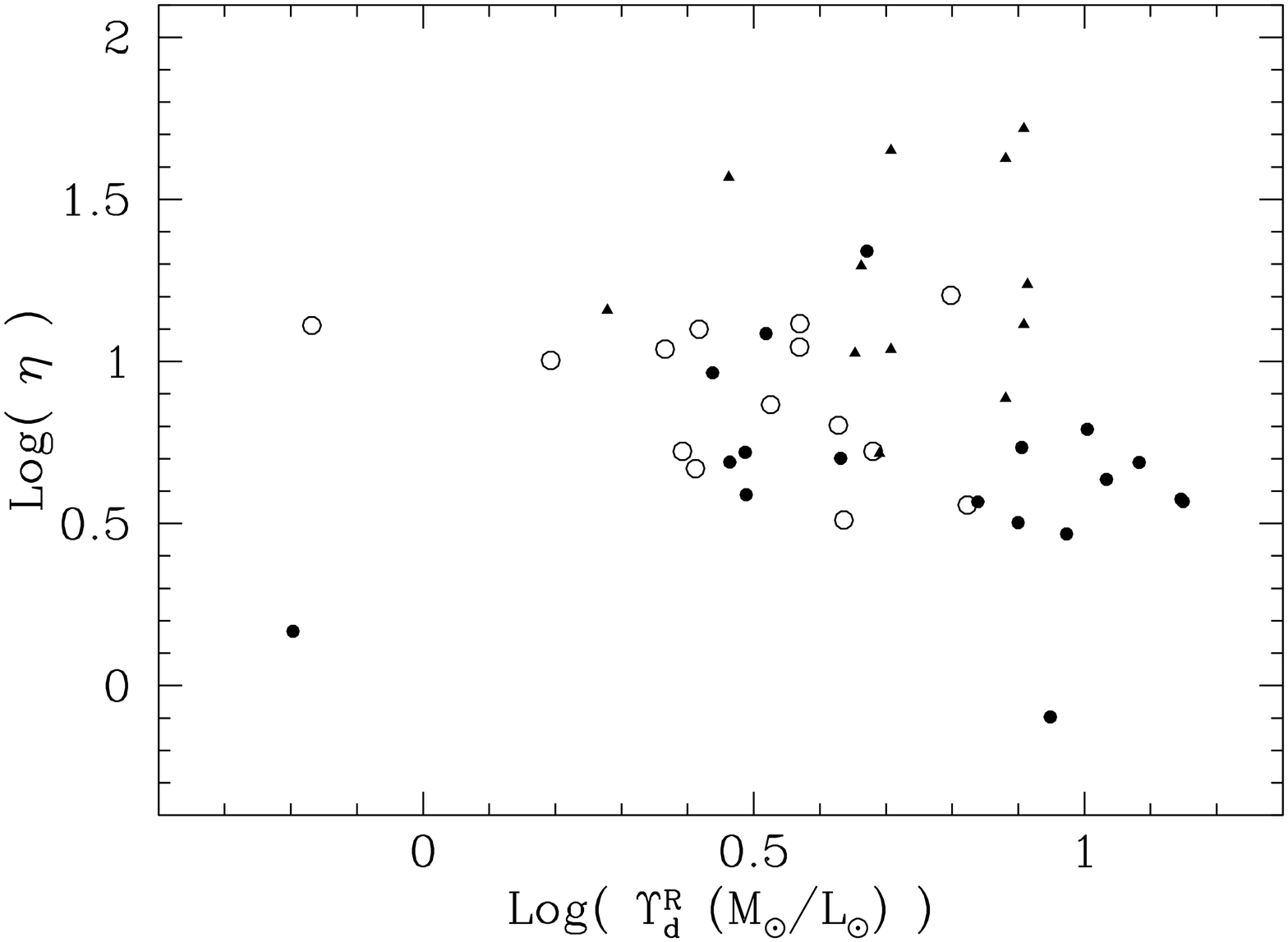}
\caption{\HI\ scaling factors $\eta$ vs. \mlstardR.  Symbol
  coding as in Figure~\ref{figlmp}.}
\label{figmgvsml}
\end{figure}

We explored the correlation of \mlstardR\ with other global
properties. Because \mlstardR\ is similar to those of maximum disk
models (see Figure 10), these correlations are similar to those found
for the maximum disk models. We find that \mlstardR\ is weakly
correlated with surface brightess (see Figure~\ref{figmgvsmu}; see
also Swaters etal 2011), but \mlstardR\ is not clearly correlated with
Hubble type or absolute magnitude (see Figures~\ref{figmgvsT}
and~\ref{figmgvsM}, respectively). In addition, as shown in
Figure~\ref{figmgvsml}, \mlstardR\ is not correlated with $\eta$,
indicating that the two scale factors are independent of each other.

Although not shown in Table~\ref{tabfits}, we have also determined
\mlstarbR, the mass-to-light ratio for the bulge for the maximum disc
and \HI\ scaling fits. Here too we find that \mlstarbR\ is similar
between the two fits, although \mlstarbR\ is approximately 10\% larger
in the \HI\ scaling fits.

\section{Discussion}
\label{secdisc}

\subsection{Baryons distribution vs. rotation curve shape}

Previous studies have already shown that in almost all galaxies the
rotation curves of the stellar discs can be scaled up to explain all
of the observed rotation curves out to two or three disc scale lengths
in spiral galaxies (e.g, Kalnajs 1983; Kent 1986; Palunas \& Williams
2000) and in dwarf and LSB galaxies (e.g., Swaters, Madore \&
Trewhella 2000; Swaters \etal\ 2011).  Thus, the detailed distribution
of the light appears to be linked to the rotation curve shape.  This
link suggests a coupling between the distributions of the total mass
and the stellar mass.  (Sancisi 2004; Swaters \etal\ 2009).

In this paper we have shown that for the vast majority of the rotation
curves of the 43 disc galaxies in our sample, ranging from early-type
spirals to late-type dwarf galaxies, the observed rotation curves can
be explained by combining and scaling the contributions of both the
stars and the \HI\ to the rotation curves.  This is consistent with
the results found previously by Bosma (1981) and confirms and extends
those of HvAS.

The link between the rotation curve shape and the galactic discs is
therefore not limited to the stars, but appears to extend to the
\HI\ discs as well. The molecular gas tends to follow the distribution
of the stars (e.g., Regan \etal\ 2001), so its contribution is
effectively included in that of the stars in our fits. Therefore, the
link between the rotation curve shape and the galactic discs seems to
extend to all the baryons in the disc.  We refer to this as the
`baryonic scaling model', because the observed rotation curves can be
explained by scaling both tracers of the baryonic mass.

The baryonic scaling model requires two free parameters, namely the
scale factors of the stellar and \HI\ discs (fits to galaxies with
bulges have three free parameters). Whereas for mass models including
dark halos there is usually a large range in acceptable fits (e.g.,
van Albada \etal\ 1985; Dutton \etal\ 2005; Swaters \etal\ 2011), for
the baryonic scaling models there is only a small range over which
acceptable fits can be found. For the baryonic scaling, there is no
degeneracy in the fits because the contributions of gas and stars to
the rotation curve peak at different radii and hence are independent
(except in a few cases, such as UGC~4325 and UGC~5414). Moreover, for
baryonic scaling to work, as was shown in Section~\ref{secmls}, and as
was also found by HvAS and Hessman \& Ziebart (2011), a (near-)maximal
disk is required.

Despite this limited freedom in the fits, the baryonic scaling can
explain the observed breadth of rotation curve shapes, from early-type
spiral galaxies with steeply rising inner rotation curves and
declining outer parts, to those of late-type dwarf galaxies with
rotation curves that rise more slowly. Even more remarkably, baryonic
scaling can also explain large-scale features in some of the observed
rotation curves (see e.g., Figure~\ref{figearly}; Broeils 1992b). We
must caution, however, that features in the rotation curves may have
different origins, such as streaming motions along spiral arms. A
one-to-one correlation between features in the observed rotation curve
and the \HI\ distribution is therefore not expected.

A global link between the baryons in a galaxy and its rotation
velocity has already been established in the baryonic Tully-Fisher
relation (e.g., McGaugh \etal\ 2000).  However, this baryonic
Tully-Fisher only links the observed baryonic mass to the amplitude of
the rotation curve. The fact that the baryonic scaling model works
well indicates that at a more local level there is also a link between
the distributions of the baryons and the rotation curve shape. Because
the rotation curve reflects the total mass distribution, the success
of the baryonic scaling model suggests a relation exists between the
distribution of the total mass and the distribution of the baryons.

\subsection{The \HI\ scale factor}

HvAS used a simple relation to estimate the \HI\ scale factor,
assuming that the mass distribution in a galaxy can be described by
two exponential discs, one for the stars and one for the \HI. Using
this, HvAS derived that the scale factor $\eta$ can be written as:
\begin{equation}
\eta = { { \Sigma_{\mathrm{stars},0} }\over{ \Sigma_{\mathrm{gas},0} } }
{{h_\mathrm{stars}}\over{h_\mathrm{gas}}},
\end{equation}
where $\Sigma_{\mathrm{stars},0}$ and $\Sigma_{\mathrm{stars},0}$ are
the central surface mass densities of the stars and the gas, and
$h_\mathrm{stars}$ and $h_\mathrm{gas}$ the exponential scale lengths
of the stars and the gas. The ratio of the scale lengths of the
stellar and \HI\ discs is roughly constant (to about a factor of two
for the sample presented here). The \HI\ central surface mass
$\Sigma_{\mathrm{gas},0}$ varies by a similar amount (e.g., Swaters
\etal\ 2002). However, $\Sigma_{\mathrm{stars},0}$ varies by a factor
of 40 across our sample, and is much higher for high surface
brightness galaxies. Thus, the scale factor $\eta$ is expected to vary
with surface brightness, as is seen in Figure~\ref{figmgvsmu}.
Because later-type galaxies, and galaxies with fainter absolute
magnitudes are more likely to have lower surface brightnesses, the
trend seen with surface brightness may also explain the trends of
$\eta$ with Hubble type and absolute magnitude.

\subsection{Baryonic scaling works}

This study clearly shows that the HI scaling previously discussed by
HvAS for a smaller sample, works well: regardless of whether the
stellar and \HI\ discs are near maximal or significantly submaximal,
the total mass density distribution inferred from rotation curves
can be reproduced by scaling the density distributions of the
stars and the \HI.  This relation between the distribution of
both the stars and the \HI\ on the one hand, and the distribution of
the total mass on the other seems to extend the idea of
a coupling between the distribution of light and total mass as
found previously (e.g., Sancisi 2004; Swaters et al. 2009) to all
baryons in the disc.

The success of the baryonic scaling model may be readily understood if
the baryons are the dominant mass component.  Indeed, in the baryonic
scaling model, the stellar disc must be maximal to explain the inner
parts of the rotation curves (see Figure~\ref{figmaxdisc}). However,
observational evidence appears to indicate that discs tend to be
submaximal (e.g., Bottema 1997; Courteau \& Rix 1999; Swaters 1999;
Kregel, van der Kruit \& de Grijs 2002; Kranz, Slyz \& Rix 2003;
Ciardullo et al. 2004; Herrmann et al. 2009; Westfall et al. 2011;
Bershady et al. 2011). As was also pointed out by Hessman \& Ziebart
(2011), the stellar light distribution can also be considered as a
proxy for the molecular gas. The maximal discs found in our baryonic
scaling model may therefore not represent maximal stellar discs, but
rather reflect the combination of the stellar disc and a molecular gas
component. Still, adding the estimated molecular gas masses to the
stellar disc masses cannot explain the high values for \mlstardR\ seen
in some of our fits. Therefore, if the baryons were the dominant mass
component, an additional cold gas component would be needed.

Several ideas have been put forward for such a cold gas component,
considering both the observed rotation curves and arguments of disc
stability (e.g., Pfenniger \& Combes 1994; Revaz et al. 2009). The
latter authors propose the presence of a 'dark gas' with a velocity
dispersion much larger than that of the HI. Additional indications
that there is more mass in the outer \HI\ discs than the \HI\ alone are
e.g. the existence of spiral structure in extended \HI\ discs (e.g.,
NGC 2915; Masset \& Bureau 2003) and the kinematics of the polar ring
galaxy NGC 4650A (Combes \& Arnaboldi 1997).

In this paper we have used two independent scale factors for the
contributions of the stars and gas to the observed rotation curve. It
is tempting to speculate whether a single scale factor could link the
combined contribution of the baryons to the observed rotation
curve. The scale factors \mlstardR\ and $\eta$ have an average range
of around ten to twenty (see Figures \ref{figmgvsT}, \ref{figmgvsmu},
and \ref{figmgvsM}). This range is too large to be explained by
variations in the stellar mass-to-light ratios or by the contribution
of molecular gas. To bring the scale factors together one has to
assume the presence of a dark baryonic component that follows the gas
distribution.

Alternatively, the success of the baryonic scaling model may imply a
coupling between the distribution of the baryons and that of the dark
matter. For example, HvAS tested the hypothesis that the dark matter
density is a scaled version of the \HI\ density, and others have
interpreted the success of the \HI\ scaling in the same way (e.g.,
Meurer \& Zheng 2011). This coupling is only straightforward to
understand if the dark matter is in the disc or in a highly flattened
halo. If the dark matter halo has a spherical shape, however, the
radial distribution of the dark matter will be different from that of
the stars and gas for the same rotation curve shape.

We have also explored baryonic scaling in the context of MOND (Milgrom
1983a,b). We created a set of MOND models by assuming an exponential
stellar disc, and using the average \HI\ profile from Swaters et
al. (2002). With this as input, we calculated the expected MOND
rotation curves for a range of different disk surface brightnesses,
assuming the acceleration parameter $a_0=1.0\times
10^{-8}$~cm~s$^{-2}$. We then fit our baryonic scaling model to these
model MOND rotation curves. The fits are generally good, and result in
\HI\ scale factors between $\sim 6$ and $\sim 10$, with lower values
found at lower surface brightnesses. The baryonic scaling model
therefore is consistent with MOND.

Finally, it is possible that the success of the baryonic scaling model
could at least in part be due to the fact that the different baryonic
components all peak at different radii, which means their individual
rotation curves can be combined to create a roughly flat rotation
curve.

\subsection{Baryonic scaling in early-type spiral galaxies}
\label{secearly}

Whether or not \HI\ scaling works by virtue of the \HI\ and stellar
rotation curves peaking at different radii can in principle be tested
by studying galaxies for which the rotation curves extend past the
peak in the \HI\ rotation curves. The early-type galaxies in our
sample make this possible, because for these galaxies the rotation
curves extend to lower \HI\ column densities (see
Section~\ref{sechiscale} and Figure~\ref{figearly}). At first sight,
as shown in Figure~\ref{figlmp}, the results may indicate that the
\HI\ scaling model starts to fail for these more extended galaxies,
because for all but one of these galaxies, the model falls below the
observed rotation curve.

However, there are several factors that may contribute to this
difference. For example, some of the \HI\ flux may be missing, either
due to the intrinsic limitations of an interferometer, or due to the
method of deriving integrated \HI\ maps, which usually includes a
thresholding step to select areas with emission. Also, it is possible
that at the low column densities observed at large radii the
\HI\ starts to become ionized, e.g., due to intergalactic ionizing
background radiation (e.g., Bochkarev \& Sunyaev 1977) or ionizing
radiation from the bright disk(e.g, Bland-Hawthorn \& Maloney 1999) in
case of warps.  Thirdly, it is possible that the kinematics of the
outermost regions do not follow the kinematics of the inner regions,
e.g., due to warps, outer spiral arms, or interactions.  Finally, as
noted in Section~\ref{secresults}, the outermost points for rotation
curves of the early-type galaxies were derived from lower-resolution
data, so it is possible that beam smearing affected the derived
rotation velocities, especially for galaxies at higher inclinations.

\section{Summary and conclusions}

We have presented mass models for a sample of 43 disc galaxies,
ranging from early-type spiral to late-type dwarf galaxies, that are
based on scaling up the stellar and \HI\ discs.  Our baryonic scaling
models fit the observed rotation curves well in the vast majority of
cases, even though the models have only two or three free parameters,
namely the scale factors, while the shapes of the rotation of the
stars and the gas are fixed. These models also reproduce some of the
detailed large-scale features of rotation curves.  In particular for
early-type spiral galaxies the models sometimes fail at large
radii. This may signal a real problem for the model, but it can also
be caused by observational effects, the analysis methods used, or
ionization of the \HI\ at low column densities. The average \HI\ scale
factor $\eta$ we find is around 8 (or 6 after correcting for the
contribution of primordial helium), although $\eta$ can vary
considerably from galaxy to galaxy.  The scale factor depends on
galaxy properties, and decreases towards lower surface brightnesses,
later Hubble types, and fainter absolute magnitudes. These results
confirm and extend those of HvAS.

A global link between the baryonic mass of a galaxy and the amplitude
of its rotation curve has already been established in the baryonic
Tully-Fisher relation. The success of the baryonic scaling model
indicates there is a more local coupling between the distribution of
the baryons and the rotation curve {\it shape}. Because the observed
rotation curve is a reflection of the distribution of the total mass,
the success of the baryonic scaling model suggests a relation between
the distributions of the baryons and that of the total mass.

\end{document}